\documentclass[12pt]{article}
\usepackage{jwstproposaltemplate} %Frankensteining things
\usepackage[numbers,compress]{natbib}
\usepackage{graphicx}
\usepackage[utf8]{inputenc}
\usepackage[
  separate-uncertainty = true,
  multi-part-units = repeat
]{siunitx}

\usepackage[compact]{titlesec}
\titlespacing*{\section}{0pt}{6pt}{3pt}
\titlespacing{\subsection}{0pt}{*0}{*0}
\titlespacing{\subsubsection}{0pt}{*0}{*0}

\usepackage{graphics,graphicx}
\usepackage[colorlinks]{hyperref}
\hypersetup{colorlinks,linkcolor={blue},citecolor={blue},urlcolor={red}} 
\usepackage{cleveref}
\usepackage[T1]{fontenc}
\usepackage[utf8]{inputenc}
\usepackage{CJK}

\begin{document}
\pagestyle{plain}
\pagenumbering{arabic}

\raggedright
\large
Heliophysics Decadal Survey 2024 White Paper \linebreak
\linebreak
\huge
Lunar Samples are Time Capsules of the Sun \linebreak
\normalsize

%\noindent \textbf{Thematic Areas:} \hspace*{60pt} $\square$ Planetary Systems \hspace*{10pt} $\square$ Star and Planet Formation \hspace*{20pt}\linebreak
%$\square$ Formation and Evolution of Compact Objects \hspace*{31pt} $\square$ Cosmology and Fundamental Physics \linebreak
%  $\square$  Stars and Stellar Evolution \hspace*{1pt} $\square$ Resolved Stellar Populations and their Environments \hspace*{40pt} \linebreak
%  $\square$    Galaxy Evolution   \hspace*{45pt} $\square$             Multi-Messenger Astronomy and Astrophysics \hspace*{65pt} \linebreak
  
\textbf{Authors:} \textit{Prabal Saxena* (NASA Goddard), Natalie Curran (Catholic University/CRESST II/NASA Goddard), and Heather Graham (NASA Goddard)}
\linebreak

\textbf{Co-Signers:} \textit{Vladimir Airapetian (American University/NASA Goddard), D.L. Blank (University of Southern Queensland), Ian Crawford (Birbeck, University of London), Ben Davidson (SpaceWeatherNews, The Mobile Observatory Project), Brett W. Denevi (Johns Hopkins University Applied Physics Laboratory), Katherine Joy (University of Manchester), James Tuttle Keane (NASA JPL), Rosemary M. Killen (NASA Goddard), Jason L McLain (NASA Goddard), Liam Morrissey (Memorial University), Noah Petro (NASA Goddard), Carle M. Pieters (Brown University), Song Tan [\begin{CJK}{UTF8}{gbsn}谭宋\end{CJK}] (Yunnan Observatories, Chinese Academy of Sciences)} 
  \linebreak

\textbf{Synopsis/Recommendation:} The history of the Sun is buried in the surface of the Moon, which presents a particularly exciting opportunity given the upcoming planned efforts for the return of humans to the Moon.  These future efforts involve long term, sustainable human exploration of the Moon and promise a return of a large mass of diverse and new types of lunar samples. The Heliophysics Decadal survey should actively embrace this coming opportunity and  facilitate cross-disciplinary efforts to unlock the secrets of the Sun held by the lunar surface.  With planned Artemis efforts that include prioritization of samples of high interest and protocols for sample handling and analysis, input into relevant solar signatures that would be most diagnostic and how best to obtain/retain them is incredibly important.  Finally, leveraging the theoretical expertise of the two communities in ways that bring them together, such as through dedicated conferences and workshops, will let the two communities help each other learn more than they could alone. 
\linebreak
\vfill 
\textbf{*Contact Email: prabal.saxena@nasa.gov}

\newpage
\par \hspace{10pt} The history of the Sun is buried in the surface of the Moon.  This is in part due to the profound effect the Sun has on bodies in the solar system \citep{1996ofm..book.....S, 2010IAUS..264..475M} - one recognized as a key goal of the previous heliophysics decadal survey ("Determine the interaction of the Sun with the solar system...") \citep{NAP13060}.  While the Sun sculpts the atmospheres, surfaces and habitability of the larger atmosphere possessing bodies in our solar system, its interactions with the Moon's surface are more direct as they have been potentially mediated by an atmosphere or a global magnetic field for only brief periods of time \citep{2017E&PSL.474..198S, NEEDHAM2017175, Weiss1246753, 2021SciA....7.7647T, 2022NatAs...6..325E}.  Consequently, interactions of both the energetic and more corpuscular (magnetized solar wind and energetic particles) outputs of the Sun are able to leave seemingly indelible signatures in the lunar surface. 

\begin{figure}[b!]
  \begin{center}
      \includegraphics[angle=0, scale=0.65]{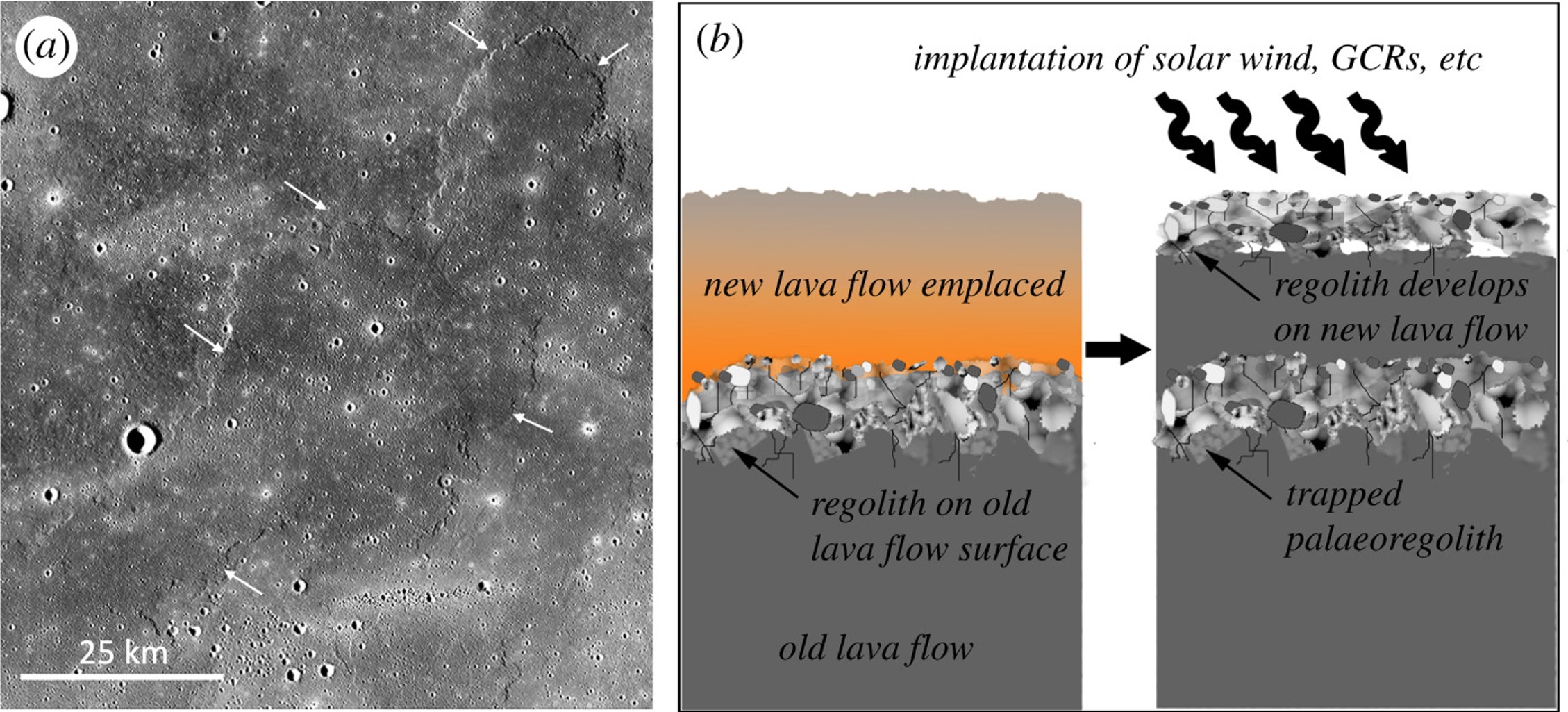}
      \caption{Examples of potential paleoregolith sites on the Moon and the process by which paleoregoliths are sealed off and may provide a valuable record of past solar interactions with the lunar surface (from \citep{2021RSPTA.37990562C}).  }
     \label{fig:paleoreg}
  \end{center}
\end{figure}

\par \hspace{10pt} A key advantage of signatures of the Sun left on the lunar surface is that there are very few processes (e.g. plate tectonics driven resurfacing, erosion, etc.) which have operated globally on the Moon and which could have wiped out such signatures.  Processes such as volcanism and impacts have modified parts of the surface, but have also done so in a way which has been able to leave many key chronologically diagnostic signatures intact.  Indeed in both cases, these process may actually serve to help highlight and constrain the chronological influence of the Sun in specific periods of time.  For example, in the case of volcanism, lava flows from episodic volcanism may have essentially sealed off layers of the regolith from further interaction with the Sun \citep{2010Icar..207..595F} that leaves these overlaid portions of `paleoregolith' as invaluable snapshots into time periods preceding the event in which they were trapped (see figure \ref{fig:paleoreg} for an example from \citep{2021RSPTA.37990562C}). Similarly, while impacts of high enough energy may in some cases modify or eliminate signatures of interest in specific regions, the impact process also drives the overturning and stratification of the regolith \citep{1974LPSC....5.2365G, COSTELLO2018327} - a process which creates a chronological record of surface interactions with the Sun with depth.

\par \hspace{10pt} The existence of potential chronologically diagnostic signatures of solar interaction with the lunar surface, particularly paleo-space weather, is especially exciting given the demonstrated ability to repeatedly actively obtain samples from the Moon.  Samples from the lunar surface have been returned to the Earth by multiple space agencies \citep{1970GeCAS...1....1S, 1971LPSC....2....1V}, beginning roughly a half century ago, and continue to this day \citep{2021Sci...374..887C}. This is true for no other body outside of Earth in the solar system, and as discussed above, lunar samples have the advantage of preserving a relatively intact record of interaction with solar and solar system processes in the past. The ability of lunar samples to explore the nature of the Sun has been recognized by scientists since the Apollo program. In 1979, a conference on `The Ancient Sun: Fossil Record in the Earth, Moon, and Meteorites' \citep{1979LPICo.390.....R} included seminal initial presentations on topics including the variation in past solar flare activity, properties of the past solar wind, changes in the solar dynamo over time and changes in GCR flux as modulated by the Sun - all deduced from examination of lunar samples \citep{1979LPICo.390...17C, 1979LPICo.390...76R, 1979LPICo.390...36F, 1979LPICo.390...12C, 1979LPICo.390...82R}.  Additional studies including some recent ones built on this work by noting the relative constancy of the solar wind in the recent past (up to 10's of millions of years) \citep{1972JGR....77..537R, 1977RSPTA.285..587C, 2018A&A...618A..96P}, but a significant number of studies, have noted evidence that the solar wind flux and consequently the Sun has not been constant farther in the past \citep{1979LPICo.390...76R, 1980E&PSL..47...34T, 1989GeCoA..53.1135B, 2006mess.book..829E}.

% \usepackage{graphicx}
\begin{table}[]
\caption{Selected Lunar Sample Proxies and Relevant Solar Inputs}
\label{tab:solarsignatures}
\resizebox{\textwidth}{!}{%
\begin{tabular}{|l|c|c|c|}
\hline
\multicolumn{1}{|c|}{\textbf{Proxy}}                              & \textbf{Meaning}                                                                                                                                    & \textbf{Solar Assumptions}                                                                                                                   & \textbf{Solar Input}                                                                                                                         \\ \hline
\begin{tabular}[c]{@{}l@{}}Cosmic Ray\\ Exposure Age\end{tabular} & \begin{tabular}[c]{@{}c@{}}Indicates exposure to cosmic rays \\ (down to several meters)\end{tabular}                                   & \begin{tabular}[c]{@{}c@{}}Assumes a constant \\ flux of GCRs and \\ SEPs over time\end{tabular}                                                          & \begin{tabular}[c]{@{}c@{}}SEP/GCR Flux/Spectrum over time,\\ Heliosphere properties that \\ modulate GCR flux/spectra\end{tabular}          \\ \hline
\begin{tabular}[c]{@{}l@{}}Particle/Fission\\ Tracks\end{tabular} & \begin{tabular}[c]{@{}c@{}}Indicates irradiation of a \\ particular sample over an\\  integrated time by\\  energetic particles\end{tabular}        & \begin{tabular}[c]{@{}c@{}}Varying assumptions about \\ the flux and plausible \\ spectra of input GCRs/SEPs\end{tabular}                                 & \begin{tabular}[c]{@{}c@{}}SEP/GCR Flux/Spectrum over time,\\ Heliosphere properties that \\ modulate GCR flux/spectra\end{tabular}          \\ \hline
Antiquity Age                                                     & \begin{tabular}[c]{@{}c@{}}Semi-quantifies formation age\\  of a breccia - the time at which a\\  soil is lithified into a breccia\end{tabular} & \begin{tabular}[c]{@{}c@{}}Dependent on isotope ratios \\ normalized to a constant\\  solar wind flux \end{tabular} & \begin{tabular}[c]{@{}c@{}}Overall SEP flux rate from Solar \\ Wind (particularly Argon), \\ Solar UV flux input for ionization\end{tabular} \\ \hline
Maturity Index                                                    & \begin{tabular}[c]{@{}c@{}}Indicates surface exposure\\  of particular sample\\  (in top mm of the surface)\end{tabular}                            & \begin{tabular}[c]{@{}c@{}}Dependent on assumption of\\  constant solar wind particle flux\end{tabular}                                                 & \begin{tabular}[c]{@{}c@{}}Solar wind flux rate - isotopes\\ of helium-3, neon-20 and argon-36\end{tabular}                                                    \\ \hline
\end{tabular}%
}
\end{table}

\par \hspace{10pt} Recent changes in understanding on how the Sun may have evolved over time have supported this past variable history and have complicated the ability to use long standing proxies used to interpret the history of and nature of lunar samples. Many of these proxies (a selection of which are listed in table \ref{tab:solarsignatures}) are dependent upon assumptions that recent research suggests are untrue, such as the assumption of constancy of energetic particle flux from the Sun over time or the constancy of the morphology of the heliosphere over time.  These processes are dependent on fundamental processes and properties of the Sun, such as rotation rate, and a number of recent studies have suggested that relaxing such assumptions leads to dramatic impacts on the atmospheres and surfaces of planets in our solar system \citep{Airapetian2016, Lammer2018} and would obviously also change some interpretation of certain lunar samples using the proxies in table \ref{tab:solarsignatures}.  

\begin{figure*}[t!]
  \centering
  \raisebox{-12pt}{\includegraphics[scale=0.40]{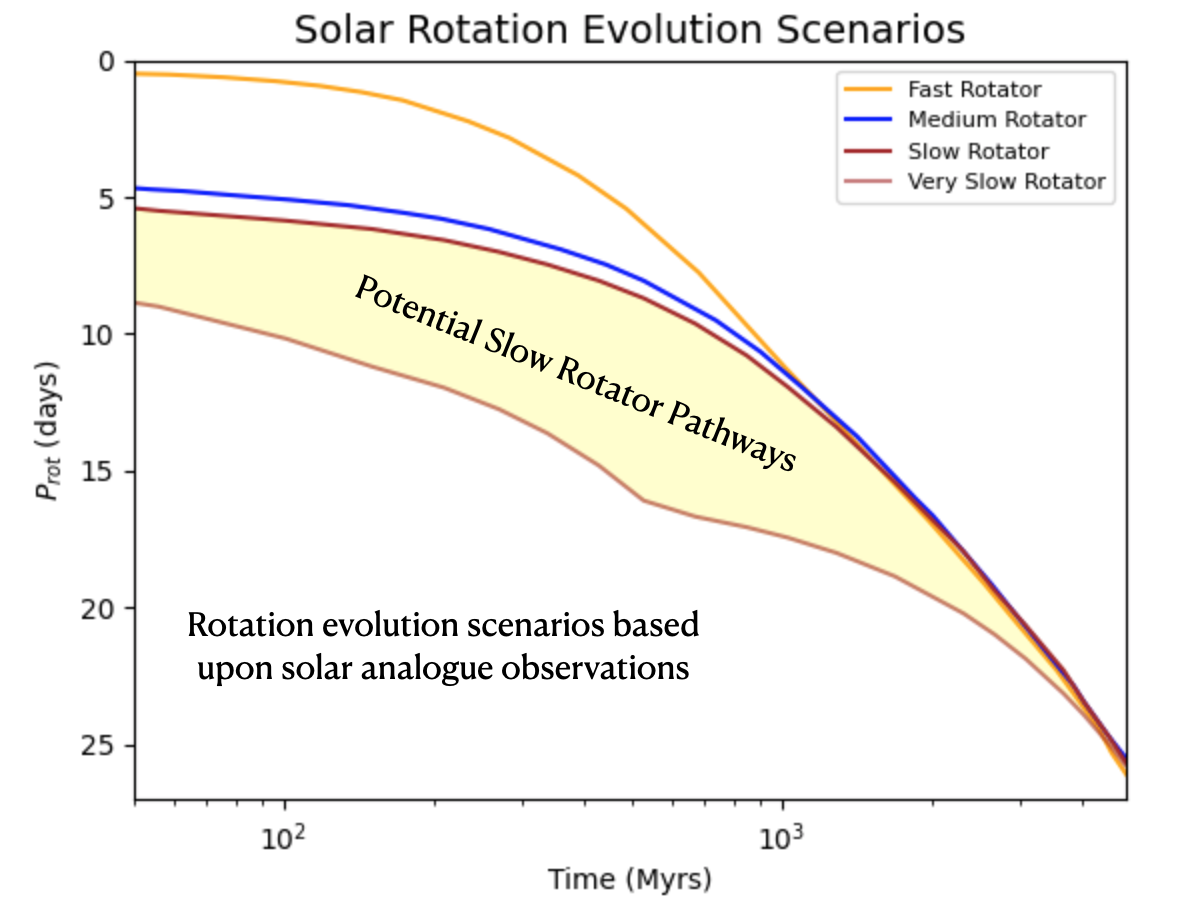}}
  \includegraphics[scale=0.30]{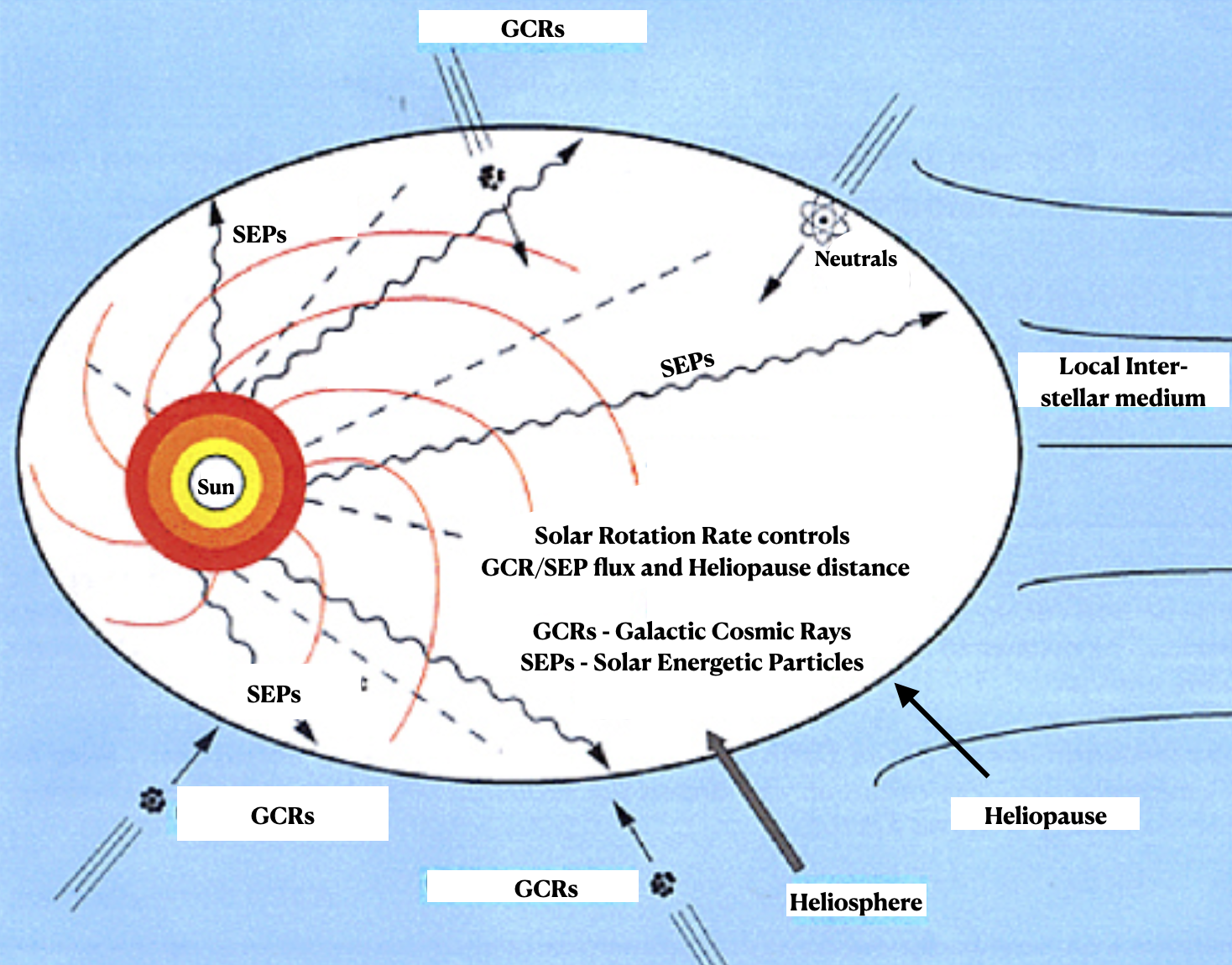}
  \caption{(Left) A plot showing potential rotation evolution scenarios for the Sun based upon solar analogue observations \citep{2013A&A...556A..36G}.  \textbf{The different potential rotation evolution pathways translate to significantly different solar activity and GCR flux.} (Right) A cartoon image of the heliosphere and different high energy particle processes relevant to estimates of cosmic ray exposure ages \citep{NAP11760} -- the changing rotation of the Sun would also influence heliosphere morphology.}
  \label{fig:solarevocartoon}
\end{figure*}

\par \hspace{10pt} Equally intriguing is that the dependence of these proxies used to examine lunar samples on these Sun related properties \textbf{may enable the use of lunar samples to put constraints on the properties of the Sun over time}.  For example, examination of solar analogue stars has suggested that the Sun may have had a range of evolutionary paths with respect to how its rotation changed over time (see figure \ref{fig:solarevocartoon} - with significant consequences for the planets in our solar system \citep{2013A&A...556A..36G, 2016A&A...587A.105A}.  The dependence of the activity of the Sun and nature of the heliosphere on these different rotation evolution paths could lead to significantly different signatures in lunar samples, particularly if there is chronological granularity provided by samples.  Indeed, initial work has attempted to place constraints on the likelihood the Sun had particular rotation states at much younger ages \citep{2019ApJ...876L..16S}. 

\par \hspace{10pt} The potential for lunar samples to hold diagnostic information regarding the nature of the Sun is especially promising giving planned return of humans to the Moon.  Many of these plans stress the importance of understanding the Sun over time using lunar samples \citep{2021LPI....52.1261W}.  These future efforts involve long term, sustainable human exploration of the Moon and promise a return of a large mass of diverse and new types of lunar samples.  This windfall of information promises to offer insight into a number of different solar system processes, and particularly relevant to studies regarding the evolution of the Sun, may enable population studies that put constraints on these processes over time. The interpretation of such signatures will require the joint expertise and efforts of planetary scientists and heliophysicists.

\par \hspace{10pt}
\fbox{%
  \parbox{\textwidth}{
    \textbf{The Heliophysics Decadal survey should actively embrace this coming opportunity and  facilitate cross-disciplinary efforts to unlock the secrets of the Sun held by the lunar surface.  With planned Artemis efforts that include prioritization of samples of high interest and protocols for sample handling and analysis, input into relevant solar signatures that would be most diagnostic and how best to obtain/retain them is incredibly important.  Finally, leveraging the theoretical expertise of the two communities in ways that bring them together, such as through dedicated conferences and workshops, will let the two communities help each other learn more than they could alone. }
    }
}

\clearpage

\bibliography{bibliography}{}
\bibliographystyle{aasjournal}

\end{document}